\documentclass{ws-procs9x6}
\usepackage{graphicx}
\newcommand {\bea}{\begin{eqnarray}}
\newcommand {\eea}{\end{eqnarray}}
\newcommand {\be}{\begin{equation}}
\newcommand {\ee}{\end{equation}}

\begin{document}

\title{Instantons and the large $N_c$ limit, A.D. 2004}

\author{T.~SCH\"AFER}

\address{Department of Physics\\
North Carolina State University\\
Raleigh, NC 27695}  

\maketitle

\abstracts{In this contribution we discuss our current understanding 
of instanton effects in the large $N_c$ limit of QCD. We argue that 
the instanton liquid can have a smooth large $N_c$ limit which is in 
agreement with scaling relations derived from Feynman diagrams. We 
also discuss certain limits of QCD, like the case of high baryon 
density, in which the Witten-Veneziano relation can be derived from 
QCD and is saturated by instantons.}

%%%%%%%%%%%%%%%%%%%%%%%%%%%%%%%%%%%%%%%%%%%%%%%%%%%%%%%%%%%%%%%%%%
\section{Introduction}
\label{sec_intro}
%%%%%%%%%%%%%%%%%%%%%%%%%%%%%%%%%%%%%%%%%%%%%%%%%%%%%%%%%%%%%%%%%%

 In the limit in which the masses of the light up, down and 
strange quarks are taken to zero, and the masses of the heavy 
quarks are taken to infinity QCD is a parameter free theory. 
This is one of the aspects of QCD that make it such a beautiful
theory, as it implies that all dimensionful numbers, like the 
masses and radii of hadrons, can be expressed in terms of a
single dimensionful quantity, $\Lambda_{QCD}$. It also implies,
however, that there is no expansion parameter that can be used 
to perform systematic calculations. 

 Many years ago 't Hooft suggested to consider the limit 
in which the number of colors, $N_c$, is large and to use
$1/N_c$ as an expansion parameter\cite{'tHooft:1973jz}. 
In order to keep the QCD scale parameter fixed we have 
to take the $N_c\to\infty$ limit with the 't Hooft parameter 
$\lambda=g^2N_c$ constant. In the large $N_c$ limit the
perturbative expansion in Feynman diagrams is replaced 
by an expansion in the genus of the two-dimensional Riemann
surface spanned by the diagrams. This result fits very well 
with the idea that the large $N_c$ limit of Yang-Mills theory 
is equivalent to a string theory. For $N=4$ SUSY Yang-Mills 
theory an explicit realization of this idea is provided by the 
AdS/CFT correspondence, but in the case of QCD the precise 
form of the string theory is not known.

\begin{figure}[t]
\includegraphics[width=11cm]{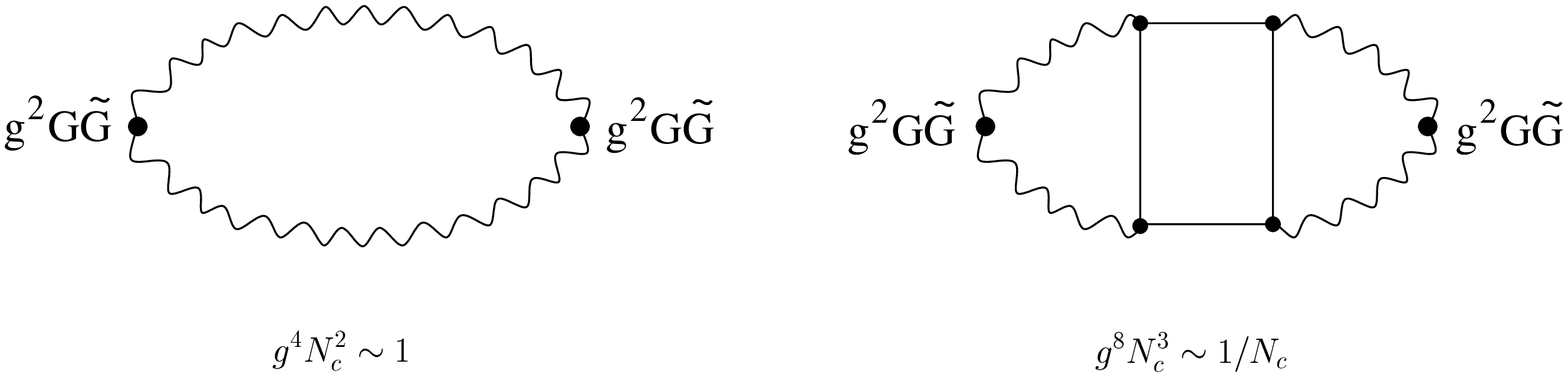}
\caption{\label{fig_scale}
The diagrams on the left and right show the large $N_c$ scaling of 
typical diagrams that contribute to the topological charge correlator
in pure gauge QCD and in QCD with light fermions.}
\end{figure}

 An interesting problem arises if we consider the fate of 
the axial anomaly in the large $N_c$ limit. For this purpose
we add a $\theta$ term 
\be 
{ L} = \frac{ig^2\theta}{32\pi^2}
               G_{\mu\nu}^a\tilde{G}_{\mu\nu}^a,
\ee
to the QCD lagrangian. The $\theta$ term is a total derivative
and in perturbation theory physics is independent of $\theta$. 
Witten suggested that non-perturbative effects generate 
$\theta$-dependence in the pure gauge theory and that the 
topological susceptibility, 
\be
\label{chi_top}
 \chi_{top} = \left.\frac{d^2E}{d\theta^2}\right|_{\theta=0},
\ee
is $O(1)$ in the large $N_c$ limit\cite{Witten:1978bc,Witten:1998uk}.
This suggestion was originally based on the fact that perturbative 
contributions to the topological charge correlator scale as $N_c^0$ 
in the large $N_c$ limit, see Fig.~\ref{fig_scale}a. Recently, Witten 
provided additional evidence for this conjecture using the AdS/CFT 
correspondence\cite{Witten:1998uk}. The scaling behavior 
$\chi_{top}\sim N_c^0$ was also observed in lattice simulations
of pure gauge QCD for\cite{Teper:2004pk} $N_c=2,\ldots,6$.

%%%%%%%%%%%%%%%%%%%%%%%%%%%%%%%%%%%%%%%%%%%%%%%%%%%%%%%%%%%%%%%%%%%
%%%%%%%%%%%%%%%%%%%%%%%%%%%%%%%%%%%%%%%%%%%%%%%%%%%%%%%%%%%%%%%%%%%

 Naive $N_c$ counting implies that the contribution of fermions
to $\chi_{top}$ is subleading in $1/N_c$. We know, however,
that there is no $\theta$-dependence in QCD with massless
fermions. This implies that the topological susceptibility
receives a contribution related to fermions that cancels the 
pure gauge result, see Fig.~\ref{fig_scale}b. Witten argued 
that this apparent contradiction can be resolved if the mass 
of the $\eta'$ meson scales as $N_c^{-1/2}$ in the large 
$N_c$ limit. Witten and Veneziano derived a relation between 
the mass of the $\eta'$ and the topological susceptibility 
in pure gauge theory\cite{Witten:1979vv,Veneziano:1979ec,Kawarabayashi:1980dp,Kawarabayashi:1980uh}
\be 
\label{WV}
  f_\pi^2 m_{\eta'}^2 = 2N_F\left.\chi_{top}\right|_{\it no\ quarks}.
\ee
Using $\chi_{top}=O(1)$ and $f_\pi^2=O(N_c)$ we observe
that indeed $m_{\eta'}^2=O(1/N_c)$. 

 The $\theta$-dependence of vacuum energy is related to
topological properties of QCD. In the semi-classical 
approximation these features can be described in terms
of instantons. Instantons are localized field configurations
that carry topological charge\cite{Belavin:fg}
\be 
 Q_{top} = \frac{g^2}{32\pi^2}\int d^4 x\,
  G_{\mu\nu}^a\tilde{G}_{\mu\nu}^a = \pm 1.
\ee
If the coupling is small then the density of instantons
scales as $\exp(-8\pi^2/g^2)$. In this limit instantons
form a dilute, weakly interacting gas. The topological
susceptibility is 
\be 
\label{diga}
\chi_{top}=\lim_{V\to\infty}
  \frac{\langle Q^2_{top}\rangle}{V}
  \simeq \frac{N}{V},
\ee
where $N=N_++N_-$ is the number of instantons and
anti-instantons and $V$ is the volume. In 1978 Witten 
pointed out that this result implies that the contribution
of instantons to the topological susceptibility scales
as $\exp(-1/g^2)\sim \exp(-N_c)$ which seems to contradict 
the assumption\cite{Witten:1978bc} $\chi_{top}=O(1)$. 

 This argument is a little oversimplified, since instantons 
in QCD come in all sizes, and only small instantons are 
exponentially suppressed. We will come back to this problem 
in Sect.~\ref{sec_liq}. Before we do so, we would like to comment 
on phenomenological consequences of equ.~(\ref{diga}). Using the 
experimental values of $f_\pi$ and $m_{\eta'}$ the Witten-Veneziano
relation implies $\chi_{top}\simeq (200\, {\rm MeV})^4$ for $N_c=3$. 
If the topological susceptibility is saturated by a dilute gas of 
instantons, this value corresponds to a density $(N/V)\simeq 1\, 
{\rm fm}^{-4}$. An estimate of the typical instanton size can be 
obtained by using the perturbative instanton size distribution 
and integrating it up to the phenomenological value of $(N/V)$. 
This leads to a value of $\rho\simeq 1/3$ fm. These two numbers 
form the basis of a successful picture of the QCD vacuum, usually
called the instanton liquid 
model\cite{Callan:1977gz,Shuryak:1981ff,Diakonov:1983hh}. The 
instanton model not only accounts for topological properties 
of the QCD vacuum, but also describes chiral symmetry breaking 
and the correlation functions of light 
hadrons\cite{Shuryak:1993,Schafer:1996wv,Schafer:2000rv}.

  Topological properties of the QCD vacuum have also been 
studied in lattice QCD. It was found that the topological 
susceptibility in pure gauge QCD is\cite{Teper:1999wp} 
$\chi_{top}\simeq (200\, {\rm MeV})^4$, as predicted by the 
Witten-Veneziano relation. It was also observed that the 
topological susceptibility is stable under cooling, and 
appears to be dominated by semi-classical configurations. 
Lattice simulations also seem to confirm the values 
of the key parameters of the instanton liquid\cite{Chu:1994}, 
$(N/V)\simeq 1\, {\rm fm}^{-4}$ and $\rho\simeq 1/3\, {\rm fm}$. 

%%%%%%%%%%%%%%%%%%%%%%%%%%%%%%%%%%%%%%%%%%%%%%%%%%%%%%%%%%%%%%%%%%
\section{Instantons and the Witten-Veneziano relation}
\label{sec_wv}
%%%%%%%%%%%%%%%%%%%%%%%%%%%%%%%%%%%%%%%%%%%%%%%%%%%%%%%%%%%%%%%%%%

 Before we discuss the $N_c$ scaling behavior we would 
like to study the mechanism for topological charge screening 
and the mass of the $\eta'$ in the instanton model. We will 
assume that instantons are small, $\rho\Lambda_{QCD}\ll 1$, 
and that the instanton liquid is dilute, $\rho^4N/V\ll 1$. 
As we shall see below, these assumptions can be rigorously
justified in the case of QCD at large baryon density. At
zero density, however, this is a model assumption.  

 The partition function of the instanton ensemble can be
written as\cite{'tHooft:1986nc,Nowak:1989at,Shuryak:1994rr}
\be
\label{Z_Q_mes}
 Z = \sum_{N_+,N_-} \frac{\mu_0^{N_++N_-}}{N_+!N_-!}
       \prod_i^{N_++N_-}d^4z_i\,\exp\left(-S_{eff}\right). 
\ee
where $\mu_0$ is the partition function of a single instanton
and the effective action is given by 
\be
 S_{eff} = i\int d^4x\, \frac{\sqrt{2N_f}}{f_\pi} \eta_0 Q
            + \int d^4x\,{ L}(\eta_0,\eta_8).
\ee
Here, the topological charge density is $Q(x)=\sum Q_i\delta(x-z_i)$ 
and ${ L}(\eta_0,\eta_8)$ is the flavor singlet sector of the 
pseudoscalar meson lagrangian 
\bea 
\label{V_eta_etap}
L &=& \frac{1}{2} \left((\partial_\mu\eta)^2 + 
   (\partial_\mu\eta_8)^2\right)  + 
  \frac{1}{2} \left( \frac{4}{3}m_K^2-\frac{1}{3}m_\pi^2\right)
 \eta_8^2  \nonumber \\
 & & \mbox{}
 +\frac{1}{2}\left(\frac{2}{3}m_K^2+\frac{1}{3}m_\pi^2\right)
 \eta_0^2 + \frac{2\sqrt{2}}{3}\left(m_\pi^2-m_K^2
 \right)\eta_0\eta_8 .
\eea
The meson lagrangian arises from bosonizing the fermionic
interaction between instantons. The pion decay constant $f_\pi$ 
is determined by the solution of a saddle point equation, see 
Sect.~IV.G. in the review article\cite{Schafer:1996wv} for 
more details. The meson masses $m_\pi$ and $m_K$ satisfy 
Gell-Mann-Oakes-Renner relations.

 The partition function equ.~(\ref{Z_Q_mes}) describes
a system of charges interacting through the exchange of 
almost massless eta mesons. The physics of this system 
is very similar to that of a Coulomb gas. We expect, in
particular, that topological charge gets screened and 
that the eta meson acquires a mass. We can show this 
explicitly by performing the sum in equ.~(\ref{Z_Q_mes}) 
and expanding the resulting cosine function to second order
in the fields. The topological charge correlator is given by
\bea 
\label{top_cor}
 \langle Q(x)Q(0)\rangle  &=& \left(\frac{N}{V}\right) \Big\{
  \delta^4(x) -\frac{2N_f}{f_\pi^2}\frac{N}{V} 
  \left[ \cos^2(\phi) D(m_{\eta'},x)\right.  \nonumber \\
 & & \mbox{}\hspace*{3.5cm} \left.
  + \sin^2(\phi) D(m_\eta,x) \right] \Big\},
\eea 
where $D(m,x)=mK_1(mx)/(4\pi^2x)$ is the euclidean space
propagator of a scalar particle. The delta-function is the 
contribution from a single instanton, while the other terms 
are the contribution of the screening cloud. The $\eta$ and 
$\eta'$ mass satisfy the Witten-Veneziano relation
\be 
\label{WV2}
  f_\pi^2\left( m_{\eta'}^2 +m_\eta^2-2m_K^2\right)
= 2N_F\left(\frac{N}{V}\right),
\ee
and $\phi$ is the $\eta-\eta'$ mixing angle. The coefficient
of the delta-function is the topological susceptibility in the 
pure gauge theory. The topological susceptibility in the 
full theory can be calculated using equ.~(\ref{top_cor}).
The result is 
\be
\label{chi_unq}
\chi_{top} =
\frac{f_\pi^2}{2N_f} m_{top}^2\left\{
  1- \frac{(\frac{4}{3} m_K^2 - \frac{1}{3} m_\pi^2)
  m_{top}^2}{(\frac{4}{3} m_K^2 - \frac{1}{3} m_\pi^2)
  m_{top}^2 +2m_K^2 m_\pi^2 - m_\pi^4}\right\},
\ee
where $m^2_{top}= m_{\eta'}^2 +m_\eta^2-2m_K^2$. The
result shows that the topological susceptibility vanishes
if any of the quark masses $m_u=m_d$ or $m_s$ is zero. 
This can be seen by using $m_\pi=0$ for $m_u=m_d=0$
and $m_\pi^2=2m_K^2$ for $m_s=0$.

%%%%%%%%%%%%%%%%%%%%%%%%%%%%%%%%%%%%%%%%%%%%%%%%%%%%%%%%%%%%%%%%%%
\begin{figure}[t]
\begin{minipage}{6cm}
\includegraphics[width=4.5cm]{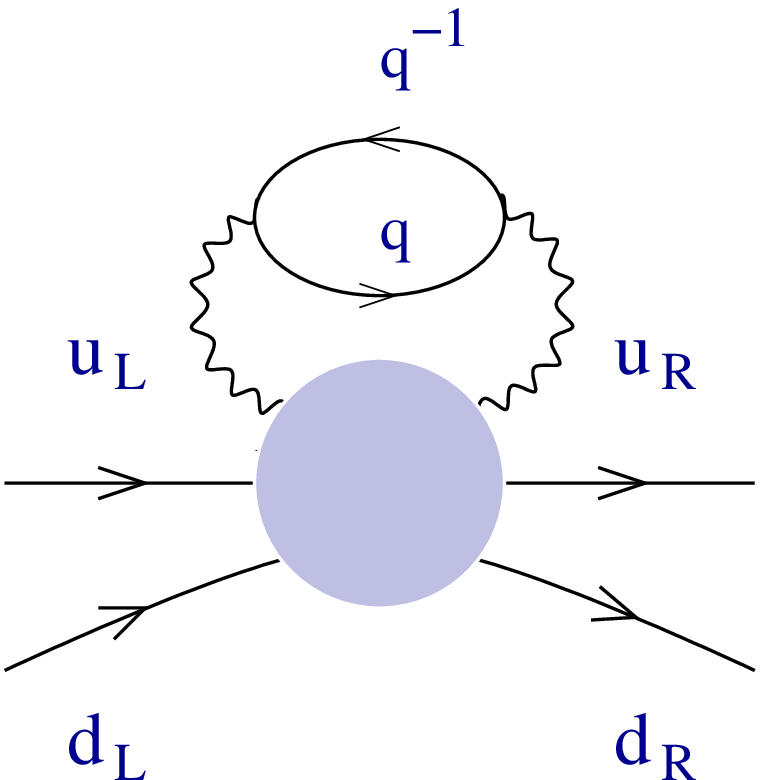}
\end{minipage}\begin{minipage}{6cm}
\includegraphics[width=5.5cm]{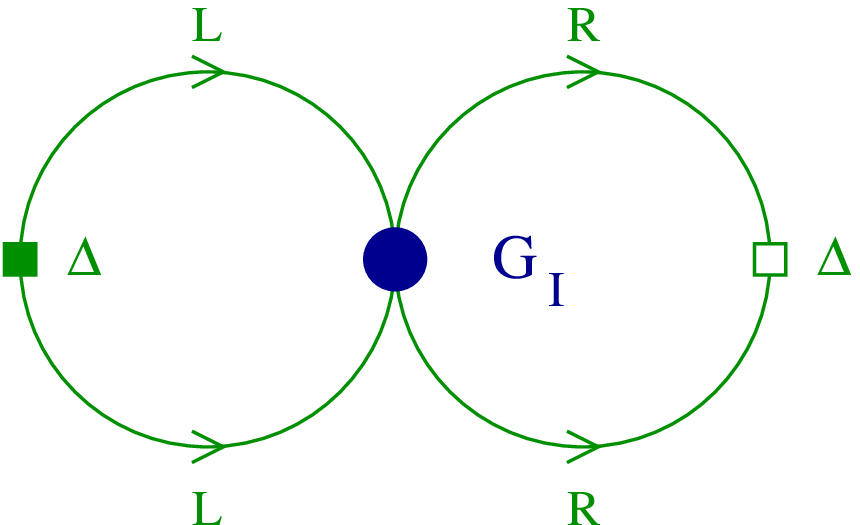}
\end{minipage}
\caption{\label{fig_imu}
The figure on the left shows the instanton induced quark 
interaction in $N_F=2$ QCD at non-zero baryon density. Scattering 
on particle-hole $qq^{-1}$ pairs leads to gauge field screening 
and suppresses large instantons. The figure on the right shows 
the instanton contribution to the vacuum energy at large baryon 
density. The squares denote insertions of the diquark condensates 
$\langle \psi\psi\rangle$ and $\langle \bar\psi\bar\psi\rangle$.}
\end{figure}
%%%%%%%%%%%%%%%%%%%%%%%%%%%%%%%%%%%%%%%%%%%%%%%%%%%%%%%%%%%%%%%%%%

%%%%%%%%%%%%%%%%%%%%%%%%%%%%%%%%%%%%%%%%%%%%%%%%%%%%%%%%%%%%%%%%%%
\section{Instantons and the Witten-Veneziano relation: Large baryon
density}
\label{sec_wv_rho}
%%%%%%%%%%%%%%%%%%%%%%%%%%%%%%%%%%%%%%%%%%%%%%%%%%%%%%%%%%%%%%%%%%

%%%%%%%%%%%%%%%%%%%%%%%%%%%%%%%%%%%%%%%%%%%%%%%%%%%%%%%%%%%%%%%%%%
\begin{figure}[t]
\begin{center}
\includegraphics[width=7cm]{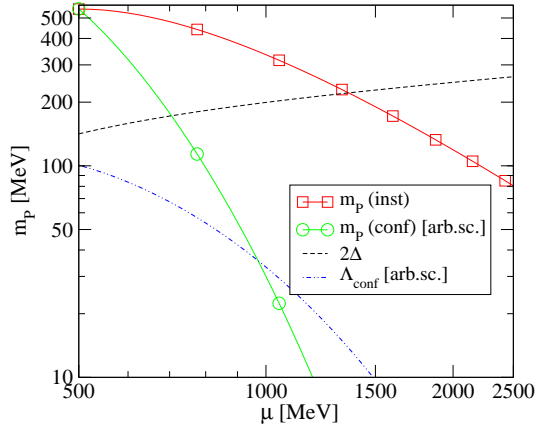}
\end{center}
\caption{\label{fig_etamu}
Flavor singlet pseudoscalar Goldstone boson mass in 
$N_c=2$ QCD. The solid line marked by squares is the result 
of the instanton calculation. For comparison, the solid line 
marked by circles shows an estimate based on the assumption 
$m_P^2f_P^2\sim\sigma^2$, where $\sigma$ is the string tension.
The overall scale of this curve is not known. We also show the 
energy gap $2\Delta$ and the confinement scale. }
\end{figure}
%%%%%%%%%%%%%%%%%%%%%%%%%%%%%%%%%%%%%%%%%%%%%%%%%%%%%%%%%%%%%%%%%%
  
  The partition function described in the last section 
provides a simple and intuitive description of the $\eta'$
prime mass and topological charge screening. The problem 
is that the theory is based on the assumption that the
instanton liquid is dilute and weakly interacting. In 
this case the screening length $l\sim m_{\eta'}^{-1}$ is 
much larger than the typical distance between charges. 
In $N_c=3$ QCD, however, the $\eta'$ is heavy and the 
screening length is very short. Instead of $\rho\ll
(N/V)^{1/4}\ll m_{\eta'}^{-1}$ we have $\rho\sim 
m_{\eta'}^{-1}<(N/V)^{1/4}$. 

 There is an interesting limit of QCD in which the 
dilute instanton liquid description can be rigorously 
justified. This is the case of QCD at large baryon density. 
It has long been known that large instantons are 
suppressed if the baryon density (or the temperature) is 
large. In the last few years it has also become clear 
that chiral symmetry remains broken at large baryon 
density\cite{Alford:1998mk}. This implies that there is a 
flavor singlet Goldstone boson, and that in the limit $m_q\to 
0$ the mass of this mode is due to the anomaly. 

  In the following we shall discuss the case of two colors 
and flavors\cite{Son:2001jm,Schafer:2002ty,Schafer:2002yy}, 
but the results can be generalized to other values of $N_c$ 
and $N_f$. At large baryon density the axial $U(1)$ symmetry
is broken by a diquark condensate $\langle\epsilon^{\alpha
\beta}\psi^\alpha_L\psi^\beta_L\rangle =-\langle\epsilon^{\alpha
\beta}\psi^\alpha_R\psi^\beta_R\rangle$. Here, $\alpha,\beta$
are spinor indices. The condensate is a color and flavor 
singlet. The $U(1)_A$ Goldstone mode corresponds to fluctuations
of the relative phase of the left and right handed condensates.
The effective lagrangian for the singlet Goldstone boson is 
\be
\label{l_nc2}
 { L} = f_P^2\left[ (\partial_0\phi)^2-v^2(\partial_i\phi)^2
 \right] - V(\phi).
\ee
The decay constant and Goldstone boson velocity are not related 
to instantons and can be determined in perturbation theory. At 
leading order the result is\cite{Son:2001jm}
\be
\label{f_nc2}
 f_P^2 = \left( \frac{\mu^2}{8\pi^2}\right),
 \hspace{1cm} v^2 = \frac{1}{3},
\ee
where $\mu$ is the baryon chemical potential. The potential 
$V(\phi)$ vanishes in perturbation theory but receives contributions 
from instantons, see Fig.~\ref{fig_imu}. We find 
\be 
\label{v_phi}
V(\phi)=-A_P\cos(\phi+\theta),
\ee
where $\theta$ is the QCD theta angle. If the 
chemical potential is big, $\mu\gg\Lambda_{QCD}$, large instantons 
are suppressed and the coefficient $A_P$ can be determined in 
perturbation theory. The result is\cite{Son:2001jm,Schafer:2002ty} 
\be
\label{A_nc2}
 A_P = C_{2,2} 6\pi^4 \left[\frac{4\pi}{g}\Delta
     \left(\frac{\mu^2}{2\pi^2}\right)\right]^2
  \left(\frac{8\pi^2}{g^2}\right)^{4}
  \left(\frac{\Lambda}{\mu}\right)^{8}\Lambda^{-2}
\ee
with 
\be
  C_{N_c,N_f} = \frac{0.466\exp(-1.679N_c)1.34^{N_f}}
    {(N_c-1)!(N_c-2)!}.
\ee
At large $\mu$ the superfluid gap $\Delta$ can also be determined 
in perturbation theory. The result 
is\cite{Son:1999uk,Schafer:1999jg,Hong:2000fh,Pisarski:2000tv}
\be 
\label{gap_nc2}
 \Delta = \frac{512\pi^4\mu}{g^{5}}
 \exp\left(-\frac{2\pi^2}{g(\mu)} -\frac{\pi^2+4}{16} \right).
\ee 
Using equ.~(\ref{f_nc2}-\ref{gap_nc2}) we can determine 
the mass of the pseudoscalar Goldstone boson. We have
\be
\label{m_P}
 m_P^2 = \frac{A_P}{2f_P^2}.
\ee
Note that we have not used the large $N_c$ limit in order to 
derive this Witten-Veneziano relation. The result is exact in 
the limit $\mu\gg\Lambda_{QCD}$ even if $N_c=2,3$. Also note 
that $A_P$ is the second derivative of the effective potential 
with respect to $\theta$ at $\theta+\phi=0$ and is equal to 
the density of instantons. The vacuum energy, however, is determined 
by minimizing $V$ with respect to $\phi$ and is independent of 
$\theta$. This implies that $\chi_{top}$ is zero, as expected 
for QCD with massless fermions. 

 Equ.~(\ref{m_P}) predicts the density dependence of the 
flavor singlet Goldstone boson mass, see Fig.~\ref{fig_etamu}.
This prediction can be tested using lattice simulations. 
We should note that formally, the prediction is only valid 
for $\mu\gg\Lambda_{QCD}$ but it is interesting to note that 
the result extrapolates to $m_P\sim 1$ GeV at zero baryon 
density. 

%%%%%%%%%%%%%%%%%%%%%%%%%%%%%%%%%%%%%%%%%%%%%%%%%%%%%%%%%%%%%%%%%%
\section{The Instanton liquid at large $N_c$}
\label{sec_liq}
%%%%%%%%%%%%%%%%%%%%%%%%%%%%%%%%%%%%%%%%%%%%%%%%%%%%%%%%%%%%%%%%%%

 In this section we describe a study of the instanton ensemble in 
QCD for different numbers of colors\cite{Schafer:2002af}. We consider 
the partition function of a system of instantons in pure gauge theory
\be
\label{Z}
 Z = \frac{1}{N_I!N_A!}\prod_I^{N_I+N_A}\int [d\Omega_I\, n(\rho_I)]
 \, \exp(-S_{int}).
\ee
Here, $\Omega_I=(z_I,\rho_I,U_I)$ are the collective coordinates of the 
instanton $I$ and  $n(\rho)$ is the semi-classical instanton distribution
function\cite{'tHooft:fv}
\bea
\label{n(rho)}
  n(\rho) &=& C_{N_c} \ \left(\frac{8\pi^2}{g^2}\right)^{2N_c} 
 \rho^{-5}\exp\left[-\frac{8\pi^2}{g(\rho)^2}\right],\\
 && C_{N_c} \;=\; \frac{0.466\exp(-1.679N_c)}
    {(N_c-1)!(N_c-2)!}\, ,\\
 && \frac{8\pi^2}{g^2(\rho)} \;=\; 
    -b\log(\rho\Lambda), \hspace{1cm} 
    b = \frac{11}{3}N_c \, . 
\eea
We have denoted the classical instanton interaction by $S_{int}$. 
If the instanton ensemble is sufficiently dilute we can approximate
the instanton interaction as a sum of two-body terms, $S_{int}=
\sum_{IJ} S_{IJ}$. For a well separated instanton-anti-instanton
pair the interaction has the dipole structure\cite{Callan:1977gz}
\be
\label{int_dip}
S_{int} = -\frac{8\pi^2}{g^2} \frac{4\rho_I^2\rho_A^2}{R^4_{IA}}
 |u|^2 \left( 1- 4\cos^2\theta \right).
\ee
Here $\rho_{I,A}$ are instanton radii and $R_{IA}$ is the 
instanton-anti-instanton separation. The relative color orientation
is characterized by a complex four-vector $u_\mu=\frac{1}{2i}{\rm tr}
(U_{IA}\tau^+_\mu)$, where $U_{IA}=U_I U_A^\dagger$ depends on 
the rigid gauge transformations that describe the color orientation
of the individual instanton and anti-instanton and $\tau^+_\mu=
(\vec\tau,-i)$. We have also defined the relative color angle 
$\cos^2\theta = |u\cdot\hat R|^2/|u|^2$. The dipole interaction
is valid if $R_{IA}^2\gg \rho_I\rho_A$. 

%%%%%%%%%%%%%%%%%%%%%%%%%%%%%%%%%%%%%%%%%%%%%%%%%%%%%%%%%%%%%%%%%%
\begin{figure}[t]
\begin{minipage}{5.9cm}
\includegraphics[width=5.4cm]{rho_nc_scal.eps}
\end{minipage}\begin{minipage}{6cm}
\includegraphics[width=6.0cm]{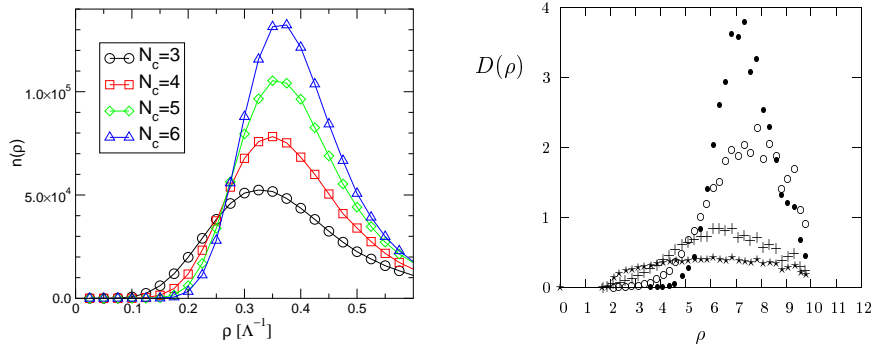}
\end{minipage}
\caption{\label{fig_size}
The figure on the left shows the instanton size 
distribution obtained from numerical simulations of the 
instanton ensemble in pure gauge QCD for different numbers
of colors. The figure on the right shows lattice results 
reported by Teper at this meeting. The $\star+\circ\bullet$ 
symbols correspond to $N_c=2,3,4,5$.}
\end{figure}
%%%%%%%%%%%%%%%%%%%%%%%%%%%%%%%%%%%%%%%%%%%%%%%%%%%%%%%%%%%%%%%%%%

   A strongly overlapping instanton-anti-instanton pair is not 
a semi-classical field configuration and we do not know how to treat 
it correctly. In practice we have chosen to deal with this problem 
by including a short range repulsive core
\be 
S_{core} = \frac{8\pi^2}{g^2} |u|^2 
    f\left(\frac{\rho_I^2\rho_J^2}{R^4_{IJ}}\right).
\ee
The precise form of the function $f(x)$ is not very important. 
What is important for the $N_c$ scaling is that the core is 
proportional to the classical action $8\pi^2/g^2$ and that 
it includes the factor $|u|^2$ which ensures that instantons
in commuting $SU(2)$ subgroups of $SU(N_c)$ do not interact. 

   The partition function equ.~(\ref{Z}) is quite complicated
and in general has to be analyzed using numerical methods. 
Before we describe numerical results we present a variational
bound. Diakonov and Petrov\cite{Diakonov:1983hh} proposed
to approximate the partition sum in terms of a variational 
single instanton distribution $\mu(\rho)$. For this ansatz
the partition function reduces to 
\be
\label{Z_var}
 Z_1 = \frac{1}{N_I!N_A!}\prod_i^{N_I+N_A}\int d\Omega_I\,
 \mu(\rho_I)= \frac{1}{N_I!N_A!}(V\mu_0)^{N_I+N_A}
\ee
where $\mu_0 = \int d\rho\,\mu(\rho)$. The exact partition
function is
\be
\label{Z-Z1}
 Z = Z_1 \langle \exp(-(S-S_1)) \rangle,
\ee
where $S$ is the full action, $S_1=\log(\mu(\rho))$ is the
variational estimate and the average $\langle .\rangle$ is
computed using the variational distribution function. The
partition function satisfies the bound
\be
\label{bound}
 Z \geq Z_1 \exp(-\langle S-S_{1} \rangle ),
\ee
which follows from convexity. The optimal distribution function 
$\mu(\rho)$ is determined from a variational principle, $(\delta
\log Z)/(\delta \mu(\rho))=0$, where $Z$ is computed from 
equ.~(\ref{bound}). One can show that the variational result
for the free energy $F=-\log(Z)/V$ provides an upper bound on 
the true free energy.  

%%%%%%%%%%%%%%%%%%%%%%%%%%%%%%%%%%%%%%%%%%%%%%%%%%%%%%%%%%%%%%%%%%
\begin{figure}[t]
\begin{center}
\includegraphics[width=7cm]{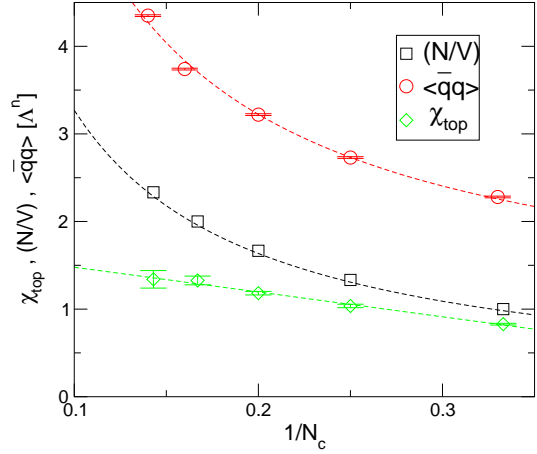}
\end{center}
\caption{\label{fig_obs}
$N_c$ dependence of the instanton density, the topological 
susceptibility and the quark condensate from numerical simulations 
of the instanton liquid in pure gauge QCD. }
\end{figure}
%%%%%%%%%%%%%%%%%%%%%%%%%%%%%%%%%%%%%%%%%%%%%%%%%%%%%%%%%%%%%%%%%%

  In order to compute the variational bound we have to determine
the average interaction $\langle S_1\rangle$. In the original
work\cite{Diakonov:1983hh} the authors used the instanton
interaction in the sum ansatz. The result is 
\bea
\label{av_int}
 \langle S_{int}\rangle  &=& \frac{8\pi^2}{g^2}
  \gamma^2\rho_I^2\rho_J^2, \hspace{1cm}
  \gamma^2 = \frac{27}{4}\frac{N_c}{N_c^2-1}\pi^2.
\eea 
For the ``dipole plus core'' interaction the result has the 
same dependence on $N_c$ but the numerical coefficient is 
different. Note that the interaction contains a factor 
$N_c/(N_c^2-1) \sim 1/N_c$ which reflects the probability that 
two random instantons overlap in color space. Since the classical 
action scales as $S_0\sim 1/g^2$ we find that the average interaction
between any two instantons is $O(1)$. Applying the variational 
principle, one finds
\bea
\label{reg_dis}
\mu(\rho) &=& n(\rho)\exp\left[ -\beta\gamma^2
 \left(\frac{N\overline{\rho^2}}{V}\right)\rho^2\right],
\eea
where $\beta=\beta(\overline{\rho})$ is the average instanton
action and $\overline{\rho^2}$ is the average size. We observe
that the single instanton distribution is cut off at large sizes
by the average instanton repulsion. The instanton density and
average size are given by
\bea
\label{dens_mfa}
\frac{N}{V} &=& \Lambda^4 \left[ C_{N_c}\beta^{2N_c} \Gamma(\nu)
(\beta\nu\gamma^2)^{-\nu/2}\right]^{\frac{2}{2+\nu}},\\
\label{rho_mfa}
\overline{\rho^2} &=& \left(\frac{\nu V}{\beta\gamma^2 N}\right)^{1/2},
\hspace{1cm}\nu = \frac{b-4}{2}.
\eea
These results imply that 
\be 
\left(\frac{N}{V}\right) \sim N_c, \hspace{1cm}
\overline{\rho}\sim N_c^0. 
\ee
Note that the instanton action is $O(N_c)$ in the large $N_c$ limit, 
i.e. instantons remain semi-classical. The total density is not 
exponentially suppressed because an exponentially large factor 
associated with different instanton embeddings cancels the 
exponential suppression from the action. Also note that the density 
scales like the number of commuting subgroups of $SU(N_c)$ and the 
instanton packing fraction $\rho^4N/(VN_c)$ is independent 
of $N_c$.

%%%%%%%%%%%%%%%%%%%%%%%%%%%%%%%%%%%%%%%%%%%%%%%%%%%%%%%%%%%%%%%%%%
\begin{figure}[t]
\begin{center}
\includegraphics[width=7cm]{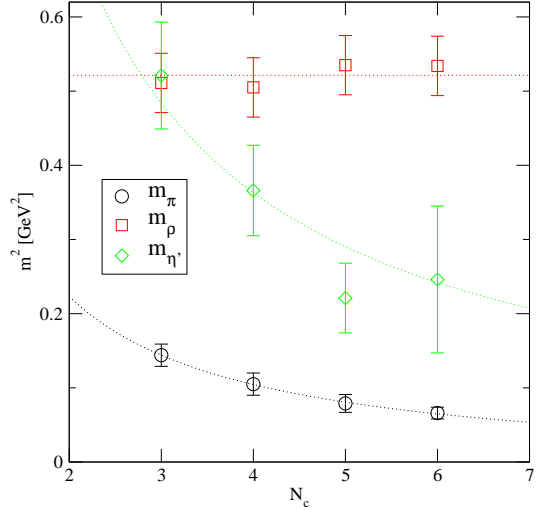}
\end{center}
\caption{\label{fig_mass}
$N_c$ dependence of the pion, the rho, and the eta prime mass 
from numerical simulations of the instanton liquid in pure gauge
QCD. The dotted lines are fits to $m_\rho^2\sim const$, $m_{\eta'}^2
\sim 1/N_c$ and $m_\pi^2\sim c_0 + c_1/N_c$.}
\end{figure}
%%%%%%%%%%%%%%%%%%%%%%%%%%%%%%%%%%%%%%%%%%%%%%%%%%%%%%%%%%%%%%%%%%

 Numerical results are shown in Figs.~\ref{fig_size}-\ref{fig_mass}.
Fig.~\ref{fig_size} shows the instanton size distribution. We 
observe that small instantons are suppressed as $N_c\to\infty$,
but there is a critical size for which the number is independent of 
$N_c$ and the total number scales as $N_c$. The results are consistent 
with the idea that the size distribution slowly approaches a delta 
function\cite{Teper:1979tq,Neuberger:1980as,Shuryak:1995pv,Munster:2000uu}. 
We also show lattice results reported by Teper at this 
meeting\cite{Teper:2004pk}. The lattice results are clearly 
consistent with our model calculation. 

 Fig.~\ref{fig_obs} shows the instanton density, the chiral
condensate and the topological susceptibility. The instanton
density and the chiral condensate scale as $N_c$. The 
topological susceptibility, on the other hand, goes to a
constant as $N_c\to\infty$. This means that $\chi_{top}$
does not satisfy the relations $\chi_{top}\sim (N/V)$ 
expected for a dilute gas of instantons. In Fig.~\ref{fig_mass}
we show the masses of the pion, the rho meson, and the 
eta prime meson. The results are consistent with the 
expectation $m^2_{\rho}\sim N_c^0$ and $m^2_{\eta'}\sim 
1/N_c$. 

%%%%%%%%%%%%%%%%%%%%%%%%%%%%%%%%%%%%%%%%%%%%%%%%%%%%%%%%%%%%%%%%%%
\section{Application: Scalar mesons and the large $N_c$ limit}
\label{sec_app}
%%%%%%%%%%%%%%%%%%%%%%%%%%%%%%%%%%%%%%%%%%%%%%%%%%%%%%%%%%%%%%%%%%

%%%%%%%%%%%%%%%%%%%%%%%%%%%%%%%%%%%%%%%%%%%%%%%%%%%%%%%%%%%%%%%%%%
\begin{figure}[t]
\begin{center}
\includegraphics[width=7cm]{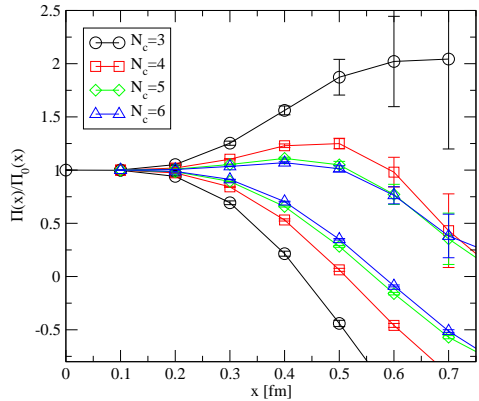}
\end{center}
\caption{\label{fig_scal}
Correlation functions in the $\sigma$ ($\bar{q}q$) and $a_0$ 
($\bar{q}\tau^aq$) channel for different numbers of colors.
The correlators are normalized to the free correlation functions
and were calculated in a pure gauge instanton ensemble. The 
fact that the $a_0$ correlator becomes negative is an artefact 
of the quenched approximation which disappear as $N_c\to\infty$.}
\end{figure}
%%%%%%%%%%%%%%%%%%%%%%%%%%%%%%%%%%%%%%%%%%%%%%%%%%%%%%%%%%%%%%%%%%

 We saw in the previous section that the instanton liquid
is consistent with the standard large $N_c$ scaling relations. 
What is maybe even more important is that instantons can be 
used to understand corrections to the leading order large
$N_c$ results. It is well known, for example, that the 
OZI rule does not work equally well in all channels. The 
OZI rule is very well satisfied in the vector channel; the 
mixing is close to ideal, and the rho and omega meson are almost 
degenerate. In the scalar and pseudoscalar channels, on the 
other hand, the OZI rule is badly violated. The eigenstates 
in the pseudoscalar sector are close to flavor, not mass,
eigenstates and the mass difference between the pion and
the eta prime meson is large. In the scalar sector we 
find a heavy iso-vector state, the $a_0$, but a light 
iso-scalar, the $\sigma$-meson, which is strongly coupled 
to $\pi\pi$ states. 

 Jaffe suggested that the unusual properties of the light 
scalar mesons could be explained by assuming a large $(qq)
(\bar{q}\bar{q})$ admixture\cite{Jaffe:1976ig,Alford:2000mm}. 
He observed that the spectrum of the flavor nonet obtained by 
coupling two anti-triplet scalar diquarks is inverted as compared 
to a standard $q\bar{q}$ nonet, and contains a light isospin 
singlet, a strange doublet, and a heavy triplet plus singlet with 
hidden strangeness. This compares very favorably to the observed 
light sigma, the strange kappa, and the heavier $a_0(980)$ and 
$f_0(980)$. It also explains why the $a_0$ and $f_0$ are 
strongly coupled to $K\bar{K}$ and $\pi\eta$.

 Instantons are important because they can account for the observed 
pattern of OZI violating effects\cite{Novikov:xj,Schafer:2000hn}.
There are no direct instanton effects in the vector channel,
and as a result OZI violation is small. In the scalar and
pseudoscalar channels, on the other hand, instantons lead
to strong flavor mixing. 

 We have recently examined instanton contributions to scalar
meson correlation functions in more detail\cite{Schafer:2003nu}.
For some earlier work on the subject we refer the reader 
to\cite{Dorokhov:1993nw}. Fig.~\ref{fig_scal} shows the 
correlation functions in the sigma ($\bar{q}q$) and $a_0$ 
($\bar{q}\tau^aq$) channel. For $N_c=3$ we find a light $\sim 
600$ MeV sigma state and a heavy $\sim 1$ GeV $a_0$ meson. When 
the number of colors is increased the light sigma state disappears 
and for $N_c=6$ the sigma mass is also in the $\sim 1$ GeV range.
We have also determined the off-diagonal sigma-pi-pi
correlation function $\langle(\bar{q}q)(0)(\bar{q}\gamma_5
\tau^aq)^2(x)\rangle$. We find that for $N_c=3$ the sigma is 
strongly coupled to two-pion states, but for $N_c>3$ the 
coupling becomes much smaller. These results are in agreement
with a study based on chiral lagrangians\cite{Harada:2003em}.

%%%%%%%%%%%%%%%%%%%%%%%%%%%%%%%%%%%%%%%%%%%%%%%%%%%%%%%%%%%%%%%%%%
\section{Supersymmetric gauge theories}
\label{sec_susy}
%%%%%%%%%%%%%%%%%%%%%%%%%%%%%%%%%%%%%%%%%%%%%%%%%%%%%%%%%%%%%%%%%%

 In Sect.~\ref{sec_liq} we argued that the large $N_c$ 
scaling behavior of the instanton contribution to QCD correlation 
functions agrees with the scaling of perturbative Feynman 
diagram. This result was based on fairly general arguments, 
but it did involve one important assumption regarding the 
effective instanton interaction that cannot be rigorously 
justified at this time. In order to study this problem in 
more detail it useful to consider supersymmetric generalizations
of QCD. 

%%%%%%%%%%%%%%%%%%%%%%%%%%%%%%%%%%%%%%%%%%%%%%%%%%%%%%%%%%%%%%%%%%%
\begin{figure}[t]
\begin{center}
\includegraphics[width=5.25cm]{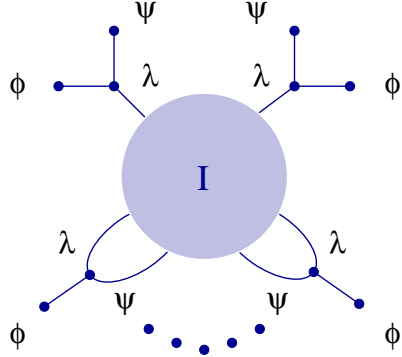}
\end{center}
\caption{\label{fig_susy}
Instanton contribution to the superpotential in SUSY gluodynamics
with $N_f=N_c-1$ quark flavors. The dots denote $(2N_c-4)$ pairs
of quark $\psi$ and gluino $\lambda$ zero modes.}
\end{figure}
%%%%%%%%%%%%%%%%%%%%%%%%%%%%%%%%%%%%%%%%%%%%%%%%%%%%%%%%%%%%%%%%%%%

 There is a significant literature on instanton effects
in supersymmetric gauge theories at large $N_c$. Instantons
in the AdS/CFT correspondence (${\it N}=4$ SUSY gauge theory)
were studied in a series of papers by Bianchi et al and 
Dorey et al\cite{Bianchi:1998nk,Dorey:1998xe,Dorey:1999pd}. 
The instanton contribution to the Seiberg-Witten superpotential 
in ${\it N}=2$ SUSY gauge theory was originally studied by 
Finnell and Pouliot\cite{Finnell:1995dr} and later generalized
to arbitrary $N_c$ by Klemm et al\cite{Klemm:1994qs} and
Douglas and Shenker\cite{Douglas:1995nw}. 

 In this contribution we would like to review some results
that are relevant to SUSY gluodynamics (${\it N}=1$ SUSY gauge 
theory). This theory is interesting because it exhibits 
confinement and a fermion bilinear condensate. It was also
recently argued that there is a new large $N_c$ limit, called
the orientifold large $N_c$ expansion, which relates the 
quark condensate in $N_f=1$ QCD to the gluino condensate 
in SUSY gluodynamics\cite{Armoni:2004uu,Armoni:2003fb}.

 SUSY gluodynamics is defined by the lagrangian 
\be 
L = -\frac{1}{4g^2}G^a_{\mu\nu}G^a_{\mu\nu}
  + \frac{i}{g^2}\lambda^{a\alpha}D_{\alpha\dot\beta}
     \lambda^{a\dot\beta},
\ee
where $\lambda$ is a Weyl fermion in the adjoint representation
and $\alpha,\dot\beta$ are spinor indices. The theory has a
$U(1)_A$ symmetry $\lambda\to e^{i\phi}\lambda$ which is 
broken by the anomaly. A discrete $Z_{2N_c}$ subgroup is 
non-anomalous. The $Z_{2N_c}$ symmetry is dynamically broken
by the gluino condensate
\be 
\label{lamlam}
\frac{1}{16\pi^2}\langle {\rm Tr}[\lambda\lambda]\rangle = 
  \Lambda^3 \exp\left(\frac{2\pi ik}{N_c}\right).
\ee
Here, $k=0,1,\ldots,N_c-1$ labels the $N_c$ different vacua
of theory and $\Lambda$ is the scale parameter defined by
\be 
\Lambda^3 = \mu^3 \frac{1}{g^2(\mu)} 
  \exp\left(-\frac{8\pi^2}{g^2(\mu)N_c}\right) .
\ee
 
 The instanton solution in SUSY gluodynamics has $4N_c$
bosonic zero modes (4 translations, 1 scale transformation
and $4N_c-5$ rigid gauge rotations) and $2N_c$ fermion 
zero modes. Reviews of the SUSY instanton calculus can be 
found in\cite{Amati:1988ft,Shifman:1999mv,Dorey:2002ik}.
The collective coordinate measure is 
\be
\frac{2^{3N_c+2}\pi^{2N_c-2} \Lambda^{3N_c}}
     {(N_c-1)!(N_c-2)!} 
 \int d^4z\, d\rho^2 (\rho^2)^{2N_c-4}\,  
 d^2\eta\,  d^2\zeta\,  d^{N_c-2}\nu\,  d^{N_c-2}\bar\nu,
\ee
where $z$ denotes the instanton position and $\rho$ is 
the instanton size. The Grassmann spinors $\eta_\alpha$ 
and $\zeta_\alpha$ parameterize the so-called supersymmetric
and superconformal zero modes, and the Grassmann numbers
$\nu,\bar{\nu}$ parameterize the superpartners of the 
rigid gauge rotations. 

 There is no direct instanton contribution to the gluino
condensate but the value of $\langle\lambda\lambda\rangle$ 
can be extracted from an indirect instanton calculation. 
The standard method consists of adding $N_f=N_c-1$ quark 
flavors to the theory, and to consider the limit in which
the squark fields have a large expectation value. In this
case instantons are small and we can reliably compute 
their contribution to the superpotential. The result 
is\cite{Affleck:1983rr,Hollowood:1999qn} 
\be 
\label{w_eff}
 W^{N_f=N_c-1,N_c}_{eff} = \frac{(\Lambda_{N_c-1,N_c})^{b_0}}
   {\det_{N_f}(Q_f\tilde{Q}_{f'})}
\ee
where $Q_f,f=1,\ldots,N_f$ are quark superfields, $b_0=3N_c
-N_f$ is the first coefficient of the beta function, and
$\Lambda_{N_f,N_c}$ is the scale parameter. Supersymmetry 
guarantees that equ.~(\ref{w_eff}) is correct even if the 
squark vev is not large. The extra quark fields can be decoupled
by sending the vev to infinity. The result is 
\be
 W^{0,N_c}_{eff} = N_c\Lambda^3_{0,N_c}
\ee
from which one can determine the gluino condensate 
equ.~(\ref{lamlam}). 

 In SUSY gluodynamics there is a direct instanton contribution 
to the $(\lambda\lambda)^{N_c}$ correlation function. The result
is independent of the relative coordinates and given 
by\cite{Hollowood:1999qn}
\be 
\label{lln}
\frac{1}{(16\pi^2)^{N_c}}
 \langle {\rm Tr}[\lambda\lambda]\ldots {
   \rm Tr}[\lambda\lambda] \rangle = 
  \frac{2^{N_c}\Lambda^{3N_c}}
       {(N_c-1)!(3N_c-1)}.
\ee
It is tempting to extract the gluino condensate from the 
$N_c$'th root of equ.~(\ref{lln}). This is sometimes called the 
strong-coupling instanton (SCI) calculation, in contrast to the 
weak-coupling instanton (WCI) result equ.~(\ref{lamlam}). The
SCI result disagrees with the WCI by a factor that scales as
$N_c$ in the large $N_c$ limit. It is not entirely clear why
the SCI calculation fails, but the problem is likely related 
to the fact that the groundstate breaks a $Z_{N_c}$ symmetry. 
This implies that the theory has to be defined carefully in 
order to pick out a unique ground state. In the WCI calculation 
the ground state is implicitly selected through the decoupling 
procedure. 

  As an alternative to the decoupling procedure one can 
define SUSY gluodynamics by compactifying the theory  
on $R^3\times S^1$. Here, both fermions and bosons obey 
periodic boundary conditions and one can show that the 
gluino condensate is independent of the size of the 
compactified dimension. In the compactified theory the gauge 
symmetry is spontaneously broken by a non-zero expectation value 
of the gauge field $A_4$. The corresponding gauge invariant order 
parameter is the Polyakov line (the holonomy) along the compact 
dimension. Instantons with a non-trivial holonomy were constructed 
by Kraan and van Baal\cite{Kraan:1998pm}. It turns out that
these objects have constituent monopoles (dyons) with 
fractional topological charge 
\be 
 S_{dyon}=\frac{8\pi^2}{g^2N_c}, \hspace{1cm}
 Q_{dyon}=\frac{1}{N_c}.
\ee
Each of these dyons has a pair of gluino zero modes that 
contribute directly to the gluino 
condensate\cite{Davies:1999uw,Diakonov:2003fi}
\be 
\frac{1}{16\pi^2}\langle {\rm Tr}[\lambda\lambda]\rangle_{dyon} = 
  \frac{1}{N_c}\Lambda^3.
\ee
Summing over the $N_c$ constituent dyons one recovers the 
WCI result for the gluino condensate. This result supports 
the old idea that in the large $N_c$ limit the relevant field 
configurations are not instantons but instantons constituents 
with fractional topological charge\cite{Diakonov:1999ae}. These 
configurations are unsuppressed because they have action 
$S\sim O(1)$. 

%%%%%%%%%%%%%%%%%%%%%%%%%%%%%%%%%%%%%%%%%%%%%%%%%%%%%%%%%%%%%%%%%%
\section{Conclusions}
\label{sec_sum}
%%%%%%%%%%%%%%%%%%%%%%%%%%%%%%%%%%%%%%%%%%%%%%%%%%%%%%%%%%%%%%%%%%

 We have argued that it is possible for the instanton liquid 
model to have a smooth large $N_c$ limit which is in agreement 
with scaling relations derived from Feynman diagrams. In this 
limit the density of instantons grows as $N_c$ whereas the 
typical instanton size remains finite. Interactions between 
instanton are important and suppress fluctuations of the 
topological charge. As a result the $U(1)_A$ anomaly is 
effectively restored even though the number of instantons grows 
with $N_c$. Using variational arguments and numerical simulations
we have shown that this scenario does not require fine tuning.
It arises naturally if the instanton ensemble is stabilized by
a classical repulsive core. In this case we obtain a picture
in which the instanton density is large but the instanton
liquid remains dilute because instantons are not strongly
overlapping in color space. Further investigations will
have to show whether this scenario is indeed correct, but 
the lattice measurements of the instanton size distribution 
reported by Teper\cite{Teper:2004pk} at this meeting are 
certainly encouraging. 

 We also emphasized that instantons provide a simple 
explanation of the observed pattern of OZI rule violating 
effects. Violations of the OZI rule are large in channels
like the scalar and pseudoscalar mesons that receive
direct instanton contributions. Instantons may also 
play a role in explaining regularities in hadron spectra
that go beyond the naive quark model, such as diquark
clustering in mesons and exotic baryons. 

 Finally we stressed that there are important lessons 
to be learned from generalizations of QCD. QCD at high 
baryon density provides a beautiful and rigorous realization 
of the instanton mechanism for generating the eta prime
mass. Supersymmetric gluodynamics is an example for a theory 
in which instantons provide the essential non-perturbative
input for the calculation of the fermion condensate. This 
calculation can now be linked, thanks to the orientifold
large $N_c$ limit, to the quark condensate in $N_f=1$ 
QCD.

\end{document}